\documentclass[11pt,twoside]{article}

\usepackage{asp2010}
\usepackage{graphicx}

\resetcounters 

\begin{document}

\resetcounters
\hskip 4in\vbox{\baselineskip12pt \hbox{FERMILAB-CONF-12-467-A}}

\title{Quantum Geometry and Interferometry}
 \author{Craig Hogan$^1$ \affil{$^1$ University of Chicago and Fermilab}}
 
\begin{abstract} 
All existing experimental results are currently interpreted using classical geometry.  However, there are  theoretical reasons to suspect that at a deeper level,  geometry emerges as an approximate macroscopic behavior of a quantum system at the Planck scale.  If directions in  emergent quantum geometry do not commute, new quantum-geometrical degrees of freedom can produce detectable macroscopic deviations from classicality:  spatially coherent, transverse position indeterminacy between any pair of world lines, with a   displacement amplitude much larger than the Planck length.   Positions of separate bodies are entangled with each other, and undergo  quantum-geometrical fluctuations that are not describable as  metric fluctuations or gravitational waves.  These  fluctuations  can either be cleanly identified or ruled out using interferometers.    A Planck-precision test of the classical coherence of space-time on a laboratory scale is now underway at Fermilab.
\end{abstract}

\section{Introduction}

Large-scale laser interferometers  have been developed  to study the dynamics of space-time with unprecedented  precision--- fractional distortions of classical geometry of less than a part in $10^{20}$, caused by gravitational waves from sources in the distant universe.   Here, I discuss the possibility  that large interferometers might measure an entirely different effect,  caused by the quantum character of  geometry itself, and originating within the space-time of the apparatus.  

At first glance this idea seems counterintuitive. New physics introduced at  small scales and high energies is usually probed by giant accelerators that collide particles at TeV energies and create interactions in attometer volumes. Quantum effects on space-time are usually thought to originate at the Planck scale, an  impossibly high energy   for accelerators.  Interferometers on the other hand appear completely classical; they measure the positions of macroscopic masses on macroscopic scales, and should seemingly be insensitive to such small scale effects. 

Yet interferometers are superb quantum measurement devices. They prepare and measure positions in states whose quantum coherence  extends over a macroscopic volume of space and time \citep{Schnabel2010,GEOnoise}. Their sensitivity currently approaches the Heisenberg quantum   limit for their size and mass.  They are also close to a physically fundamental  threshold of precision: a power spectral density for position noise given by the Planck time, where deviations from classicality might be expected.  In these respects, interferometers are uniquely well suited  to measure or quantitatively constrain tiny quantum deviations from 
classical features of geometry, such as  separation of large and small scales, the independence of  positions in different directions, and  the principle of locality.

\section{Quantum geometry} 

It is hard not to take the classicality of macroscopic geometry  for granted. The  idea of a position in space is the first physics we all learn as small children, before we even think of space as part of physics.
But from the perspective of  quantum physics, the large scale classical coherence of space is a deep mystery.

The standard operational theory of the physical world is assembled from  two distinct pieces. The first is geometry: a classical, dynamical  space-time, that is the stage for everything else. The second includes all the forms of quantum matter and energy that move and transform in time and space as particles and fields.  The two pieces are spliced together in a  way that is itself classical, and self-consistent on large scales: the quantum character of the stuff in the energy-momentum tensor is ignored for gravitational purposes, and quantum particles and fields move about within a classically determinate  space-time.

This way of joining of the quantum world with geometry works  well to explain every experimental result in physics.  On the other hand, just because a theory is consistent and successful in a certain range of applications does not mean that it is complete, or correct in all circumstances.  Indeed,  there are good theoretical reasons to suspect  that at  a deeper level,  geometry has a quantum character:

\begin{itemize}
\item
The expansion of the universe is observed to be accelerating. This behavior is controlled by the gravitation of the vacuum, which is simply an arbitrarily chosen constant in standard theory.  Its explanation lies outside the standard paradigm of fields propagating in classical space-time \citep{RevModPhys.61.1,Cohen:1998zx,Frieman:2008sn}. 
\item
Thought experiments that include curved space-times, such as black holes,  show that the dynamics of gravity and space-time can be interpreted as a statistical behavior, like thermodynamics \citep{Jacobson:1995ab,PhysRevD.81.124040,Verlinde:2010hp}. That is, the equations of Newton and Einstein can be derived on the basis of statistical principles from the behavior of new, as yet unknown quantum degrees of freedom.   The number of fundamental degrees of freedom appears to be holographic \citep{RevModPhys.74.825}:  information about the state of a causally connected space-time volume can be encoded with Planck information density on its two-dimensional boundary.   This nonlocality and limited information content cannot be reconciled in a fundamental theory with only classical geometry and quantum fields. 
Similarly, thought experiments that include black holes and fields show effects like Hawking evaporation--- essentially, a conversion of geometry into particles. For quantum principles to hold, the geometry must have holographic quantum degrees of freedom.
\item
The fundamental mathematical structures of quantum mechanics and classical geometry are entirely different, and their splicing is not controlled by any well defined mathematical limiting procedure \citep{RevModPhys.29.255,PhysRev.109.571}. 
In quantum mechanics, a position is a property of an interaction, and is described by an operator; in classical geometry,  a position is a property of an event, and is described by a real vector. There is no physical way to compare positions of classically-defined events. The standard way of splicing these two different mathematical concepts together is self-consistent at low energies, but since it assumes  classical behavior, it  excludes  quantum-geometrical effects {\it a priori}. 
\item
Similarly, the  geometrical concept of spatial localization, at the heart of classical geometry, is not a property of  reality. Quantum physics is not consistent with ``local realism'',  as  now demonstrated by many  real-world experiments \citep{Ma2012,RevModPhys.75.715}.  Although no experiment has yet directly revealed a quantum property of geometry itself, we also do not know how to reconcile experiments with the idea that classical geometry can be ``real'', since it can only be measured with quantum processes that fundamentally do not happen in a definite time or place.  Quantum mechanical nonlocality is sometimes described as paradoxical,  but from the point of view of quantum mechanics, the apparent classical coherence of space at  large separation may be the deeper mystery.
\item
Beyond the Planck scale, a dynamical classical  geometry is no longer consistent with quantum mechanical matter.  A quantum particle confined to a sub-Planckian volume in three dimensions has a mass exceeding that of a black hole in that volume, impossible according to relativity;  conversely, a black hole with mass below the Planck scale has a quantum position indeterminacy larger than its Schwarzschild radius, so the geometry must be indeterminate. 

\end{itemize}

Many promising ideas for unifying classical and quantum descriptions have been pursued over the last century.  Decades of mathematical literature document consistent progress in quantum theories that include gravity, such as string theory, matrix theory, loop quantum gravity, and noncommutative geometry.  They  allow a consistent description of physics at the Planck scale and beyond \citep{Hossenfelder:2012jw}.  They  also display explicit holographic dualities  in curved space-times; for example, a conformal quantum field theory on the boundary of an Anti-De Sitter space also describes a quantum theory of matter and gravity in the higher-dimensional bulk. 
On the other hand, no microscopic quantum theory yet gives a clear account of the emergence  of a macroscopic, nearly-classical, nearly-flat spacetime--- that is, a realistic laboratory setting--- so the connection of these ideas with classical geometry has not been tested experimentally.  

The  approach taken here does not derive from  gravity or quantized fields, or  from any particular fundamental microscopic theory.   Instead we  use general principles of special relativity and quantum mechanics to
 directly estimate possible new macroscopic  effects  of Planckian quantum geometry, if positions in different directions do not commute.
These arguments suggest that
interferometers  may detect effects  of quantum geometry on the positions of massive bodies.

\section{Emergent Space-time}

One promising,  general approach to quantum geometry is to suppose that  classical space-time is ``emergent''.  The general idea is that classical notions of spatial direction,  position,  and locality may arise only as approximations,  in a macroscopic limit. On small length scales, the system becomes less classical and ``more quantum''.  At the Planck scale,   geometrical  states become fully indeterminate quantum systems.  

To make this idea work in practice, the classical limit should reconcile standard physics with  hints of quantum geometry just identified, such as the holographic behavior of gravitational states.  Macroscopic symmetries of space and time, such as Lorentz invariance, should be derived rather than assumed.
Ideally, some new predictions for realistic experiments might also emerge, that could confirm that these ideas have something to do with the real world.

According to one idea \citep{Banks:2011av}, the Hilbert space of a  space-time, together with all the matter in it,  is defined  in relation to  a particular timelike world line.   In the emerged space-time, an interval on the world line defines a causal diamond--- a region in the intersection of the future light cone of the initial point, and the past light cone of the final point.   The state  associated with an interval that lasts for $N$ Planck times is represented by an $N\times N$ matrix that represents everything happening within a causal diamond.  

This construction is  holographic: the number of degrees of freedom is the area of the covariantly-defined 2D bounding surface in Planck units. By construction,  it is consistent with causality and general covariance.  Since it is built around a particular world line, it is not manifestly  consistent with full Poincare invariance; whether or not this is a problem, is a quantitative issue to be settled by experiment. 
It has been suggested that physics based on an emergent space-time could provide a natural setting to explain both inflation and cosmic acceleration \citep{Padmanabhan:2012gx,Banks:2011qf}. 
 
Emergent space-time is a useful framework to discuss new effects of quantum geometry on the positions of bodies in nearly flat space.  
It allows us to contemplate new violations of classicality, such as position operators in different directions that do not commute with each other.  Although quantum geometry originates in Planck scale physics, in an emergent space-time its effects need not be confined to Planck scale frequencies or scales; it can  be spatially nonlocal, shared coherently by many particles; and it can produce distinctive, observable, entangled fluctuations of macroscopic positions.

\section{Noncommuting macroscopic quantum geometrical position operators}

A position is described by an operator, that operates on a state describing a system.
Position operators are not unique, but can represent various ways of preparing and measuring a quantum state.  Some operators correspond to conventional position operators; for example, the position operator for a particle operates on a subset of the system, corresponding to that particle, and correlates it with another subsystem representing a  measurement apparatus. Indeed it is common practice to approximate systems of interest as idealized isolated subsystems, and ignore other degrees of freedom. Such a subsystem is  conventionally idealized as an isolated state or prepared system, but it is really part of a larger state that includes that of the geometry it inhabits. 

A similar procedure can be followed for  new quantum-geometric modes. We can ignore {\it all} the standard quantum degrees of freedom,  and  write down a quantum theory  of operators that  represent only new collective geometrical position degrees of freedom in an emergent system, that are shared by many particles.  This program is less ambitious than most approaches to Planckian physics, since it does not attempt to formulate a fundamental, microscopic theory.  The main constraint is  that  the overall behavior agrees sufficiently well with classical space-time position to agree with experiments.

Consider the mean position of some massive collection of particles in a compact region of space, which we call a ``body''.   Suppose that the position of  a  body in each direction $\mu$ is a quantum observable, represented by a self-adjoint operator $x_\mu$.  The commutators of these operators represent the quantum deviations of a massive body from a classical trajectory. The body itself is assumed to be massive enough that we  ignore the conventional position operators--- the usual quantum effects associated with its motion.

To describe the  quantum degrees of freedom of the geometry, posit the following  commutators relating positions  in different directions \citep{Hogan:2012ne}:
\begin{equation}\label{covariant}
[x_\mu,x_\nu]= \bar x^\kappa \bar U^\lambda \epsilon_{\mu\nu\kappa\lambda}  i\ell_P,
\end{equation}
where indices $\mu,\nu,\kappa,\lambda$ run from 0 to 3 with the usual summation convention,  
$\bar x^\kappa$ denotes the expectation value of the position, $\bar U^\lambda\equiv \partial \bar x^\lambda/c\partial \tau$  the dimensionless expected 4-velocity of the body, $\tau$ the proper time,  $\epsilon_{\mu\nu\lambda\kappa}$ the Levi-Civita antisymmetric 4-tensor, and $\ell_P$  a parameter with the dimensions of  length.

In the limit $\ell_P\rightarrow 0$, the commutator vanishes, so that  positions in different spatial directions behave independently and classically.
 It is interesting to ask what happens if the scale  $\ell_P$ is not zero, in particular if it is  of order the Planck length,  $\ell_P\approx ct_P\equiv \sqrt{\hbar G/c^3}= 1.616\times 10^{-35}$ meters.  With this choice the number of the geometrical degrees of freedom approximately agrees with  holographic entropy bounds for gravitating systems. 

One virtue of equation (\ref{covariant}) is that it is  manifestly covariant:  the two sides transform in the same way under the homogeneous Lorentz group, as a direct product of vectors.  The algebra of the quantum  position operators  respects the transformation properties of corresponding coordinates in  an emergent classical Minkowski space-time, in a limit where the operators are interpreted as the usual space-time coordinates. 
The theory itself  thus defines no preferred direction  in space. These operators are thus plausible candidates for classical positions in the macroscopic limit.

Indeed the form of departures of positions from classical behavior--- the commutator--- depends on classical position and 4-velocity in a  way that is determined by the need for covariance.
The quantum commutator of two vectors requires two antisymmetric indices that must be matched by indices  on the right side.
 Thus we require a nonvanishing antisymmetric tensor, which in four dimensions has  four indices, $\epsilon_{\mu\nu\lambda\kappa}$.  Two of its antisymmetric indices match those of the noncommuting positions.  The other two must contract with two different vectors to avoid vanishing. The unique geometrically defined options are the 4-velocity and position of the body being measured. 
 
On the other hand,   Eq. (\ref{covariant}) is not {\em invariant}. The commutator does depend on the position and 4-velocity of the body being measured, or equivalently, on the origin and rest frame of the coordinate system.  We interpret this to mean that the  commutator describes a quantum relationship between world lines that depends on  their relative positions and velocities,  but not on any other properties of the bodies being compared.  In Eq. (\ref{covariant}), the quantum-geometrical position state of a body is defined in  relation to a particular world line,  the origin of the coordinates. 

These attributes are expected  if   quantum geometry describes a relationship between timelike trajectories. Unlike a classical metric defined independently of any observer, the state of a quantum geometry is shaped by a choice of world-line, so as noted above, it cannot obey Poincare invariance. The reference world line is defined in this instance by the coordinate system.  

In the rest frame of the body being measured,
the  4-velocity  is  $\bar U^\lambda= (1, 0, 0, 0)$ so the  non-vanishing terms of Eq. (\ref{covariant}) are those multiplied by $\epsilon_{\mu\nu\kappa\lambda}$ with $\lambda=0$.  The remaining terms  describe a noncommutative geometry in three dimensions:
\begin{equation}\label{3Dcommute}
[x_i,x_j]= \bar x^k \epsilon_{ijk} i\ell_P,
\end{equation}
 where indices $i,j,k$ now run from 1 to 3, and  the operators $x_i$ correspond to positions at a single time, in the rest frame of the body. Eq. (\ref{3Dcommute})   describes a quantum-geometrical relationship between positions of two trajectories (or  massive bodies) that have  expected proper 3-separation
$\bar x^k$,  and whose world-lines have the same expected 4-velocity.

\section{Quantum geometrical position uncertainty} 

As usual in quantum mechanics, the operators represent observables, and they operate on states that represent physical systems. In this case, the quantum system describes the  geometry that relates the trajectories, which is usually assumed to be classical.  Also as usual, if we think of the  position state as  represented by a wave function, rather than a matrix, we can estimate the quantum indeterminacy in position.  The wave function in this case is not invariant, but depends on the positions and velocities of the trajectories whose relationship it describes. In particular, the complementarity of position in different directions depends on the separation vector. It depends only on the mean position and velocity, consistent with describing a collective degree of freedom of many particles, that  is, a massive body or bodies.

 The quantum commutator leads to an uncertainty relation  in the usual way, although the conjugate variables are now positions in different directions, instead of  familiar examples such as position and momentum. 
In the rest frame,  the uncertainty relations for  a body at  position $\bar x^k$  are 
\begin{equation}\label{3Duncertainty}
\Delta x_i\Delta x_j \ge   |\bar x^k \epsilon_{ijk} | \ell_P /2,
\end{equation}
where $\Delta x_i= \langle |x_i- \bar x_i |^2\rangle^{1/2}$ represents the spread of the wave function in each direction, and $\langle\rangle$ denotes an average over the wave function.

Remarkably,   the wave functions of position in the directions transverse to separation $\bar x^k$ between trajectories  show a  quantum-geometrical uncertainty  that  actually increases with  $|\bar x|$. For trajectories with macroscopic separation, this new uncertainty is much larger than a Planck length. 

One consequence is that the notion of spatial locality emerges self-consistently, over durations much longer than a Planck time. The quantum-geometrical uncertainty within a small region of space-time scales like the duration $\tau'$ of a causal sub-diamond, $\Delta x_i\Delta x_j \approx  c\tau'  \ell_P$.  Everything in that region coherently shares a larger quantum-geometrical  deviation from classical position, relative to a   distant  world line with $\tau>>\tau'$. 

 Classical  space-time emerges  as an excellent approximation to describe positions and trajectories with separations much larger than the Planck length. Consider the angular uncertainty, from Eq. (\ref{3Duncertainty}), in direction to a body  on the 3-axis, with an expected position $(0,0, \bar x_3)$:
\begin{equation}
\Delta \theta_1\Delta \theta_2 \ge   \ell_P /2 |\bar x_3|,
\end{equation}
where $\Delta\theta_1=\Delta x_1/|\bar x_3|$  and $\Delta\theta_2=\Delta x_2/|\bar x_3|$.  
For separations on any experimentally accessible scale, this  deviation from classicality is fractionally negligible.
However, as separations approach  the Planck scale,  directions become mostly  indeterminate.   The classical approximation  breaks down, consistent with the idea of a space-time emerging from a Planckian quantum system.

The transverse position  uncertainty can be related to  holography by counting degrees of freedom. The number of independent  positions in the radial direction is  the diamond duration, $c\tau/\ell_P\approx |\bar x|/\ell_P\approx N$. The  number of independent transverse  states in both transverse directions is  about  $
|\bar x|^2/\Delta x_i\Delta x_j\approx |\bar x|/\ell_P\approx N$, so the product in all three directions is $\approx N^2$, as required for position states that give rise to holographic gravity.

\section{Measurements, fluctuations, and classicality}


Uncertainty (as in Eq. \ref{3Duncertainty}) refers to the width of a wave function, but of course this function is not  measured. It has a width  at a particular time in the rest frame, but in a time series measurement, the uncertainty manifests as fluctuations or noise. The material in each  patch normal to a separation vector from  the reference world line appears to undergo a coherent  transverse random walk  of about a Planck length per Planck time relative to the immediately interior patch.  Because the effect is transverse, it cannot be detected by measurements between just two world lines but requires at least three world lines, and a spatially extended measurement in two directions.  As discussed below, these requirements can be met in a suitably configured interferometer.

These quantum-geometrical fluctuations have no direct relation to vacuum fluctuations of Planck-scale modes of quantum fields, or of the metric.  They are due to a quantum indeterminacy in the spatial relationships of timelike trajectories of large aggregations of particles, rather than a zero point oscillation of a field mode.  The  new degrees of freedom that originate in the noncommutative geometry have normal modes that are not plane waves. They combine wildly different longitudinal and transverse  scales. 

In a typical laboratory experiment, on the scale of a few meters, $N$ is the of the order of $10^{36}$.  The  equivalent speed of the spatially-coherent transverse geometrical fluctuation  is about $N^{-1/2}c$, or about one centimeter per year--- a tiny speed generally associated with long, slow processes, such as motions of the earth's crust. Here however the coherence time $\tau$ for the  fluctuations is the light travel time across a laboratory, typically tens of nanoseconds, and the total (transverse) excursion on that timescale is of the order of ten attometers.  Averaged over longer durations, the fluctuations around classical positions are even smaller. This tiny departure from classicality  would have escaped detection up to now.

Recall that the entire state of a causal diamond of duration $\tau= N t_P$ is represented by an $N\times N$ matrix.
Typically,  states corresponding to particles  are  of the order of $N^{1/2}\times N^{1/2}$ in size, with $N$ total degrees of freedom. That is far less than the $\approx N^3$ degrees of freedom in a field theory in the same volume with a Planckian cutoff.  Physically, the reason for the reduction is that  quantum geometry entangles field modes in different directions: they are no longer independent.

On the other hand , even  the space-time within a single elementary-particle collision in a collider such as the Tevatron or the LHC, on the TeV scale, comprises $N\approx (m_Pc^2/{\rm TeV})\approx 10^{16}$. This number  is still so large that  quantum-geometrical effects on the phase space of particle interactions would have escaped notice at the attainable levels of experimental precision in colliders. For this purpose, even an attometer is macroscopic compared with the Planck scale.

If the mass $m$ of a body is less than the Planck mass, $m<m_P\equiv \hbar/c^2t_P=2.176\times 10^{-5}$ g, the standard Heisenberg  uncertainty \citep{PhysRevLett.45.75} for the variance in a body's position difference measured at two times separated by a duration $\tau$,
\begin{equation}
\Delta x^2\equiv \langle (x(t)-x(t+\tau))^2\rangle \ge 2\hbar\tau/m,
\end{equation}
is greater than the quantum-geometrical position uncertainty at separation $c\tau$. 
 Quantum-geometrical uncertainty is therefore  negligible on the mass scale of elementary particles ($\approx $ TeV $\approx 10^{-16} m_Pc^2$), which helps to explain why classical space-time is such a good approximation for systems involving small numbers of  particles, and why standard theory agrees so well with precision tests in microscopic experiments.

Quantum-geometrical position entanglement thus only becomes significant, compared to standard quantum mechanics, in large aggregations of particles. Of course, the effect is not generally noticed since  correlations, and much larger displacements, arise  from  all the usual interactions between particles in a typical massive body. However, in a special, carefully prepared  system such as an interferometer, these tiny, purely geometrical  position displacements of free massive bodies  can be   decoupled from other environmental factors, and measured.

\section{Response of interferometers to quantum-geometrical uncertainty}

The positional quantum states of bodies in quantum geometry  possess a kind of  nonlocal coherence not describable by states  of standard quantum theory in classical space-time.  In the standard view, the position of a massive body is an average over many particles; the macroscopic, very low  frequency components of particle motion are highly correlated, and  reduce to only the three classical positional degrees of freedom for the body as a whole. Here,  an additional coherent entanglement of geometrical position states creates a new  correlation in the mean positions of otherwise separate bodies--- an in-common, coherent quantum-geometrical deviation from their classical trajectories.

 In a Michelson interferometer, the  normal modes of  photon fields are shaped by the boundary conditions, particularly the beam splitter, into  combinations of plane waves in two directions  \citep{Caves:1981hw}.  The signal at the dark port of the interferometer corresponds to a position-difference operator that
coherently entangles the  position states of three massive bodies in two directions, separated by the arm length. In a quantum geometry, positions in the two directions are not independent, and  quantum-geometrical position entanglement on this scale  affects the signal.

For a simple Michelson interferometer, the response of a signal to  quantum-geometrical uncertainty resembles a Planckian random walk of the beamsplitter position up to durations given by twice the arm length, $\tau= 2L/c$.
A more precise estimate of the predicted displacement power  spectrum is \citep{Hogan:2010zs}
\begin{equation}\label{displacementspectrum}
\tilde\Xi(f)= {4c^2t_P\over \pi (2\pi f)^2}[1-\cos (f/f_c)], \qquad f_c\equiv c/4\pi L.
\end{equation}
This quantity gives the mean square  displacement in measured arm length difference, per frequency interval. 

The spectrum at  frequencies above $f_c$ oscillates with a decreasing envelope that scales like $\tilde\Xi(f)\propto f^{-2}$. At  frequencies much higher than $f_c$, the mean square fluctuation in a frequency band $\Delta f$ goes like $\tilde\Xi(f)\Delta f \propto (\Delta f/ f)(c^2 t_P/f)$.  This result is independent of $L$,  as it must be since it  results from a universal noise that depends only on the Planck time, and shows  increasing total variance in position at low frequency, as reflected in the uncertainty relation (Eq. \ref{3Duncertainty}).
The apparatus size acts as a cutoff: quantum-geometrical fluctuations from long duration modes ($c\tau>2L$) do not add noise to the signal, so that the noise spectrum at  frequencies  below $f_c$ approaches a constant.
In addition,  the mean square displacement averaged over a time $\tau$ much longer than $2L/c$ is
$ \approx (4ct_PL/\pi) (2L/c\tau)$,
showing that the effect in a given spatial volume decreases in a time averaged experiment; again, over long durations, everything acts more like a classical system.
Since the frequency spectrum of the displacement  flattens off at frequencies  below the inverse system size,  detection of the fluctuations is optimized with a time resolution comparable to the system size.

If two interferometers are near each other--- that is, if they probe mostly the same space-time volume--- their geometrical position states are entangled, even if they have no physical connection apart from proximity.  By correlating their signals, one can measure the entanglement of geometrical position states. Depending on the configuration of two interferometers, the cross-correlated signal is a measure of both the amplitude of the geometrical fluctuations, and of their entanglement.

The predicted noise spectrum includes no parameters apart from  known scales: the size of the apparatus, and the Planck time. It includes distinctive features such as zeros that signify its origin in the relative positions of the optical elements.  The spectral shape, its amplitude, and its  spatial correlation are all measurable quantities.  The theory thus offers a clean target for experimental test.

\section{Real Interferometers}

If  quantum-geometrical noise exists, it contributes to noise in gravitational wave detectors. However, its effects are different from gravitational waves, so the response depends on details of the interferometer optical layout.

 At  LISA frequencies, in the millihertz band, quantum-geometrical noise  will be hidden beneath   a confusion-limited background from many sources of gravitational waves.  Future detectors (like the Big Bang Observer) that resolve the confusion background from binaries in the ~0.1 to 1 Hz band will be affected by quantum-geometrical noise, if it exists.
 
 The most sensitive operating detector in the band from 0.1 to 1 kilohertz, LIGO, is  not much affected by quantum-geometrical noise,  because its optical design is relatively insensitive to transverse displacements: most of the response of its signal to gravitational waves is generated in arm cavities.  At  frequencies in its detection band, which are far below the inverse light-travel time,  its sensitivity is dominated by longitudinal displacements that are free of quantum-geometrical noise.  GEO600 is the currently operating detector most sensitive to the new effect, and indeed   already operates close to the predicted Planckian  noise level (the low frequency limit of Eq. \ref{displacementspectrum}).  However, it is  not configured to isolate the particular  signatures of  quantum-geometrical noise that distinguish it from other noise sources, so  it  is not optimized to make a definitive test.

The Fermilab Holometer is an experiment designed  specifically to  detect the Planckian quantum-geometrical noise, if it exists, and to rule it out, if it does not. The basic layout is a pair of 40-meter Michelson interferometers in close proximity.  Corresponding optical elements of the two machines are within  a meter or two of each other, so their signals probe almost the same instantaneous space-time volume; their causal diamonds mostly overlap. Position fluctuations are measured at  high  frequency, up to tens of MHz, to resolve the predicted transfer function (Eq. \ref{displacementspectrum}).   High frequency operation also allows a simpler design than gravitational wave detectors; in particular, mechanical isolation from the environment is much simpler. 

  Theory predicts that the correlated signal should reveal a new source of continuum noise with a spectrum close to Eq. (\ref{displacementspectrum}), with a critical frequency $f_c= 6\times 10^5$ Hz and a first zero at $f= c/80{\rm m}= 3.75$ MHz.  The cross-correlation offers several advantages:   integration over time  reduces the relative importance of other noise sources, such as the dominant photon shot noise; alternative configurations allow response to the quantum geometry fluctuations to be ``turned off''; and  specific diagnostics can be investigated in the time domain, such as vanishing correlation beyond a lag $\tau=2L/c$.  

If  quantum-geometrical noise can be measured, its  properties will convey detailed information about  the  relationship between classical and quantum geometry, and the statistical interpretation of gravity.    
If the predicted Planck-amplitude noise does not exist, then it might be said that we have merely ruled out a particular interpretation of emergent space-time. However, the result will stand as a solid constraint on the nearly-classical coherence of space-time with Planckian sensitivity, that must be obeyed by any future theory that seeks to explain the origin of classical geometry from first principles.

\acknowledgments
This work was supported by the Department of Energy at Fermilab under Contract No. DE-AC02-07CH11359.

\bibliography{HoganBibli}
\bibliographystyle{asp2010}

 \end{document}